\newcommand{\CLASSINPUTbottomtextmargin}{1.01in}
\documentclass[conference,10pt]{IEEEtranTCOM}
\IEEEoverridecommandlockouts

\normalsize

\usepackage{amsmath,amsthm,amssymb,graphicx,tikz, bbold,bbm}

\usepackage{amsfonts}
\usepackage{etoolbox}
\usepackage{mathtools}
\let\labelindent\relax
\usepackage{enumitem}
\usepackage{capt-of}
\usepackage{textcomp}
\usepackage{etoolbox}
\usepackage{float}
\usepackage[caption=false]{subfig} 
\usepackage{array}
\usepackage{multirow}
\newcolumntype{P}[1]{>{\centering\arraybackslash}p{#1}}
\newcolumntype{M}[1]{>{\centering\arraybackslash}m{#1}}

\newcounter{pfxc}[section]

\def\BibTeX{{\rm B\kern-.05em{\sc i\kern-.025em b}\kern-.08em T\kern-.1667em\lower.7ex\hbox{E}\kern-.125emX}}

\preto\subequations{\ifhmode\unskip\fi}
\newcommand{\rvec}[1]{{\boldsymbol{\mathbf{\mathsf{\MakeLowercase{#1}}}}}}

\newcommand{\rs}[1]{{\boldsymbol{\mathbf{{#1}}}}}
\newcommand{\s}[1]{#1}

\newcommand{\ddt}{\frac{\partial}{\partial \tau} }

\begin{document}
\title{Elements of disinformation theory: cyber engagement via increasing adversary information consumption}

\author{Travis C. Cuvelier,  Sean Ha, Maretta Morovitz 
\thanks{T. Cuvelier, S. Ha, and M. Morovitz are with the MITRE Corporation, McLean, VA, 22102 USA e-mails: tcuvelier@mitre.org, seanha@mitre.org,  mmorovitz@mitre.org. This work was supported by the MITRE Internal Research and Development Program. Copyright 2024 The MITRE Corporation. Approved for public release, case: 24-2326.}
}

\maketitle

\begin{abstract}
We consider the case where an adversary is conducting a surveillance campaign against a networked control system (NCS), and take the perspective of a defender/control system operator who has successfully isolated the cyber intruder. To better understand the adversary's intentions and to drive up their operating costs, the defender directs the adversary towards a ``honeypot" that emulates a real control system and without actual connections to a physical plant. We propose a strategy for adversary engagement within the ``honey" control system to increase the adversary's costs of information processing. We assume that, based on an understanding of the adversary's control theoretic goals, cyber threat intelligence (CTI) provides the defender knowledge of the adversary's preferences for information acquisition. We use this knowledge to spoof sensor readings to maximize the amount of information the adversary consumes while making it (information theoretically) difficult for the adversary to detect that they are being spoofed. We discuss the case of imperfect versus perfect threat intelligence and perform a numerical comparison. 
\end{abstract}

%\begin{IEEEkeywords}
 %Control systems, control with communication constraints, LQG, source coding.
%\end{IEEEkeywords}

% \fi
\IEEEpeerreviewmaketitle

\section{Introduction}
To drive up cyber adversaries' costs and better learn their tactics, techniques, and procedures (TTPs), network defenders can opt to \textit{engage} intruders rather than rely on traditional passive defenses and blacklisting. Motivated by recent threats targeting critical infrastructure and industrial control systems (ICS) \cite{voltTyphoon},  we propose an engagement technique specialized to the domain of networked control systems. We consider a scenario where an adversary, conducting reconnaissance, would like to convey structured measurements from a cyberphysical system for further processing at a remote location. We assume that the network defenders, employing deception operations, have successfully diverted the adversary towards a honeypot, namely an emulated cyberphysical sensor output mechanism (dubbed a honey programmable logic controller (PLC)). This could be accomplished by a combination of deception techniques based on Software Defined Networking \cite{ha2023dataplane} and classical lures and honeytokens. As the physical process observed by the honey PLC is emulated, defenders can freely perturb or spoof the sensor outputs. Under an assumption that the defender has knowledge of how the adversary quantizes the sensor feeds for remote consumption, we propose to optimize the spoofing to trade off between increasing the adversary's required information consumption at the remote site with the enhancing the adversary's  
ability to detect deception locally.

For common control and estimation tasks, there are established techniques on how to quantize and encode sensory data to effectively achieve control theoretic goals with a limited communication bitrate and/or sensing overhead. The motivation of that prior art includes reducing both communication resource consumption and information processing requirements as well as extending battery lifetimes. These approaches broadly fall under the purview of ``rationally inattentive" control and sensing. Essentially, task performance requirements dictate the quantizer designs, and quantizer outputs are encoded via standard lossless data compression. We propose that adversaries targeting NCS will be motivated to use communication-efficient encodings when feeding sensor information back to remote locations. Adversaries have an incentive to minimize their use of the defender's network to remain undetected, and have a further incentive to minimize the amount of data they will be required to process offsite. Consider, for example, an attack against a bank where the adversary's goal is to remotely monitor when a vault is unlocked. If, through local scripting on the defender's network, the adversary can deploy an automated agent to check the status of the vault periodically, they might choose to communicate (e.g. send a ping) only when an ``unlock" occurs (as the vault is typically locked). Assume instead that the adversary views a \textit{fictitious} set of vault control and associated sensor readings that are under the direction of defenders. Our engagement technique essentially suggests that the defenders spoof the sensor feeds to make it appear that more ``vault unlock" events occur (increasing the number of pings to the remote site), while controlling for the adversary's wariness of spoofing.

\subsection{Organization}
In the remainder of this section we review some related literature and outline our notation. We proceed with a detailed problem definition in Section \ref{sec:pf}, which leads to an optimization to identify optimal sensor-spoofing distributions. We solve this optimization for some cases of practical interest in Section \ref{sec:optimization}. A numerical example is used to build intuition in Section \ref{sec:numerical}. We conclude in Section \ref{sec:conclusion}. 
\subsection{Related work:}
There is a vast literature of networked control with communication and/sensing constraints. In particular, there is significant work on task-guided, real-time, lossy data compression for common control tasks. Put succinctly, control performance replaces the notion of ``distortion" from classical rate-distortion theory and additional constraints are introduced to account for real-time requirements. The general setup consists of a ``sensor" that periodically samples and quantizes sensory data which is losslessly encoded into packets. The packets are conveyed to (remote) ``controllers" or ``estimators"  who make time-sensitive decisions. For many relevant control/estimation tasks, near optimal encoding, quantization, and decision methods are known. An early tutorial review on quantizer design for control applications can be found in \cite{yuksel} and a general model for control and estimation with informational constraints is found in \cite{shafieepoorfard106rationally}. Our present work is aligned with approaches that permit packet lengths to vary over time. This is often practical necessity with unstable systems, and permits the use of ``event-based" communication (where an encoder can choose to remain silent). Such quantization and encoding schemes for linear-quadratic-Gaussian control and Gauss-Markov tracking can be found in \cite{kostinaTradeoff,tanakaISIT,ourJSAIT,cuvelier2023algorithms}. More recent work considering event-based encoding includes \cite{guo2021optimal}. Presently, we will assume that the adversary uses communication-efficient encodings of sensor feeds and that their choice of quantization function is accessible (in the black box sense) to the defenders. The prior art illustrates connections between an adversary's control theoretic goals and the structure of their quantization and encoding methods. 

There is a wealth of literature on networked control focused on approaches to secure cyberphysical systems. A recent review can be found in \cite{dibaji2019systems}. Recent attacks on control systems are categorized according to their ability to disclose information, distrupt operations, and their required a priori knowledge of the system itself \cite{Teixeira2012Taxonomy}. The review covers work on robistifying control systems against cyberattack-induced failures and proposed frameworks for data security and privacy. Some active methods for intrusion detection and system resilience are discussed; we describe a subset below. 

Many proposed approaches to active cyber defense use tools familiar to the control community (whether or not specialized to cyberphysical systems). Analytical approaches include Stackelberg security games (cf. e.g. \cite{Alpcan2003game,Sanjab2017game}) and modeling the deployment of security countermeasures in terms of partially observable Markov decision processes (POMDPs) (e.g. \cite{miehling2018POMDP,hammar2020finding}). These approaches are somewhat limited in that they require explicit or implicit access to models that govern both adversary and defender behavior in the abstract. One needs to, e.g., quantify reward structures that motivate  adversary behavior, and establish a dynamic model for how this behavior affects systems of interest. This seems challenging in practice. While the engagement approach we propose in this work requires the defender to have knowledge of how  adversaries quantize sensory information, we believe this to be a more concrete notion of ``threat intelligence". We anticipate that given an understanding of an adversary's control theoretic goals, the prior art on rationally inattentive control can be used to identify quantization approaches an adversary might use.

There is comparatively less work on how to engage and deceive cyber adversaries in ICS or cyberphysical systems. One notable exception is a baiting approach introduced in \cite{flamholz2019baiting}.  \cite{flamholz2019baiting} proposes that defenders physically disturb the plant (or, equivalently, introduce adversaries to false plant models) to make it easier to detect stealthy attacks on state estimators. Also relevant is recent work on deceptive feedback control \cite{Ornik2018Deception,karabag2022deception,patil2023simulator}. In \cite{Ornik2018Deception}, a framework is proposed from the perspective of an agent who seeks to maximize a control theoretic reward that depends on the beliefs of an observer. An augmented POMDP optimization problem is formulated and attacked with standard techniques. While \cite{Ornik2018Deception} begins to address the difficulty of establishing POMDP models for an adversary's (in this case, the observer's) beliefs, the need for a dynamic model presents a significant impediment. Our present work incorporates several ideas from deception in supervisory control as introduced in \cite{karabag2022deception,patil2023simulator}. In designing an approach to sensor spoofing, we control for the adversary's ability to detect spoofing. We take a similar tack to \cite{karabag2022deception,patil2023simulator} and use a Kullback-Leibler (KL) divergence penalty motivated by hypothesis testing to control for adversary suspicion. 

\subsection{Notation:} We use boldface $\rvec{y}$ to denote random variables, and script letters $\mathcal{C}$ to denote sets. For sequences,  $\{y_{t}\}$ denotes $y_{0},y_{1},... , y_{\infty}$ and, e.g. $y_{1:5}$ denotes $y_{1},y_{2},\dots,y_{5}$. Integrals are of the Lebesgue type. We occasionally rely on measure-theoretic probability notation, and use equivalent notation according to our present context. We generally use  $\Omega$ (further annotated with subscripts or superscripts) to denote (probability) spaces and $\mu$ (often with annotations) to denote (probability) measures over a canonical sigma algebra $\sigma(\Omega)$. The $T$ fold product of a the sigma algebra $\sigma(\Omega)$ is $\sigma(\Omega)^{T}$, and the set of all probability measures defined over elements of a sigma algebra is, e.g. $\mathcal{P}(\sigma(\Omega))$. For $\mathcal{C}\in \sigma(\Omega)$ let $\mathbbm{1}_{x\in\mathcal{C}}:\Sigma\rightarrow\mathbb{R}$ be an indicator function with $\mathbbm{1}_{x\in\mathcal{C}} =1$ if $x\in\mathcal{C}$ and $\mathbbm{1}_{x\in\mathcal{C}} = 0$ otherwise. If $\rvec{y}$ is drawn from the probability measure $\mu$ we write $\rvec{y}\sim \mu$. Then, we have, for a measurable set $\mathcal{C}\subset \Omega$, $\mathbb{P}[\rvec{y}\in\mathcal{C}] = \mu[\mathcal{C}]$. $\mu^{1} \ll \mu^{2}$ denotes that $\mu^{1}$ is absolutely continuous with respect to  $\mu^{2}$, e.g. that for all $\mathcal{C}$, $\mu^{2}\left[\mathcal{C}\right]=0$ implies  $\mu^{1}\left[\mathcal{C}\right]=0$. If  $\mu^{1} \ll \mu^{2}$, then there exists a Radon-Nikodym (RN) derivative $\frac{d\mu^1}{d\mu^2}:\Omega \rightarrow \mathbb{R}_{+}$ such that if $\mathcal{A}$ a measurable subset of $\Omega$ then $\mu^{1}[\mathcal{A}] = \int_{x\in\mathcal{A}} \frac{d\mu^{1}}{d\mu^{2}}(x)d\mu^{2}(x)$. If $\rvec{y}\sim\mu$ is real valued and $\mu$ is absolutely continuous with respect to Lebesgue measure, we say $\rvec{y}$ is a continuous random variable with a probability density function (PDF) $f_{\rvec{y}}$ such that $\int_{x\in\mathcal{C}}f_{\rvec{y}}(x)dx = \mu[\mathcal{C}]$.  We will abuse notation when referring to  ``marginalized" densities and RN derivatives. For example, if $\{\rvec{y}_{i}\}$ is a sequence on $\Omega$ drawn from $\mu_{\rvec{y}}$ and  $\mathcal{A}_{k}\subset \Omega^{k+1}$ we will write  $\mu_{\rvec{y}}[\mathcal{A}_{k}\times \Omega^{T-k}] =\mu_{\rvec{y}_{0:k}}[\mathcal{A}_{k}]$, or, equivalently $\mathbb{P}\left[ \rvec{y}_{0:k} \in \mathcal{A}_{k} \right] = \mu_{\rvec{y}_{0:k}}[\mathcal{A}_{k}]$. Let $\{0,1\}^{*}$ denote the set of all binary strings of finite length. For $x\in \{0,1\}^{*}$, let $\ell(x)$ denote the length of $x$, in bits. We include the empty string $\emptyset\in \{0,1\}^{*}$ such that $\ell(\emptyset) = 0$.

\section{System model and problem formulation} \label{sec:pf}
\begin{figure}[h]
    \centering
    \includegraphics[width=.9\linewidth]{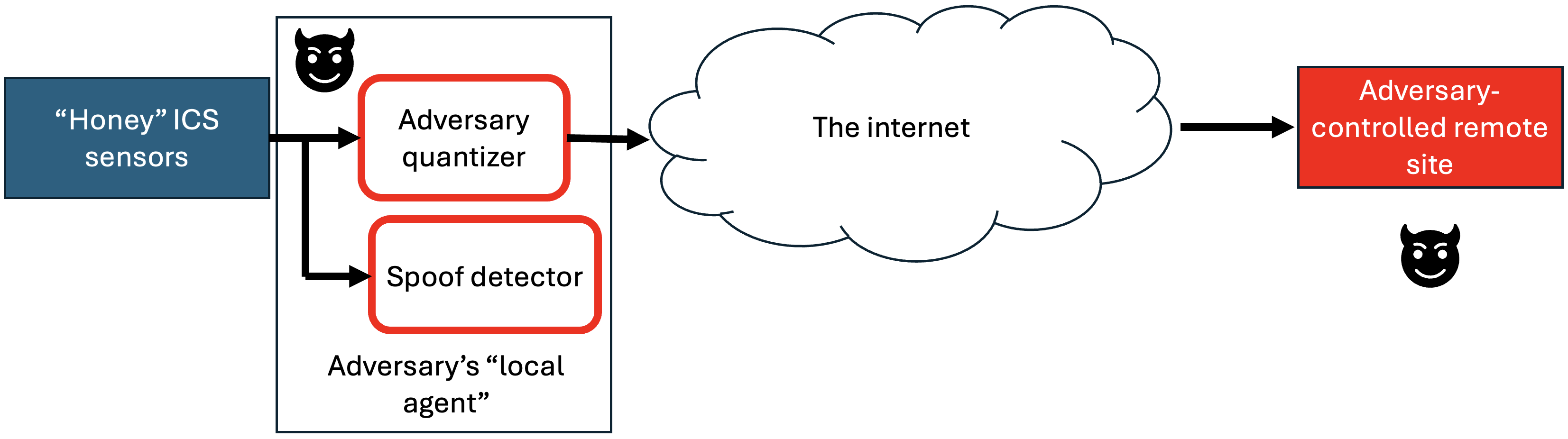}
    \caption{The scenario of interest. We assume that we have successfully lured an adversary into engaging with a ``honey PLC", believing its sensor outputs to be authentic. The adversary's malware carries out reconnaissance and processing ``locally" on the \textit{defender's} network. The malware is tasked with selecting and quantizing a subset of the available sensor feeds and reporting to a remote C2 server for further analysis. We refer to the malware running on the defender's network as the adversary's ``local agent", and the remote C2 server as the ``remote site". 
    }
    \label{fig:scenario}
    \vspace{-3mm}
\end{figure} 
We consider the scenario in Figure \ref{fig:scenario}, where defenders have successfully lured an adversary into accessing a ``honey" sensor feed in an ICS. We assume that at the present phase of the attack, the adversary's goal is to perform reconnaissance of the controlled plant. While the honey sensor readings appear to mimic the output of sensors physically connected to the plant, they are rather controlled by the defense team. We call this ``sensor spoofing". We assume that the adversary uses malware run on the defender's network to select, encode, and relay a subset of the sensor readings for processing at a remote command and control (C2) server. We refer to the malware as the ``local agent", the C2 server the ``remote site", and the selection and discretization of sensor data as \textit{quantization}. We view the  adversary as pursuing a line of attack and is conducting reconnaissance to inform said attack - these sensor readings being a critical component of the reconnaissance -  and assume that he will be satisfied to proceed to the next phase of the attack after receiving some number of messages from his local counterpart. For example, the adversary might be trying to accomplish real-time parameter estimation, and the attack might proceed when the estimate error at the remote site drops below some threshold. We assume that the adversary is suspicious and that the local agent runs a hypothesis test to decide whether or not the sensor readings are being spoofed (which would indicate that the defenders are employing deception). If the local agent detects a spoof, he will sever the connection to the remote C2 server and the adversary will either pivot or proceed to a subsequent phase of the attack. We propose that the defender can drive up the adversaries' expenses and unnecessarily prolong their dwell time through active engagement by \textit{requiring the adversaries to consume more information than they would otherwise prefer}. We now formulate this mathematically.

We assume that the sensor feeds are a discrete-time stochastic process $\{\rvec{y}_{t}\}$ on a probability space $(\Omega,\sigma(\Omega),\mu)$. The space $\Omega$ can be discrete or continuous. As the local agent communicates with the remote site via an external network, e.g. the Internet, the local agent must first quantize the sensory data, and encode the quantized measurements (\textit{quantizations}) into packets. We use the word ``quantization" in a general sense; we consider the space of sensor readings $\Omega$ to be partitioned into a discrete, countable set of events of interest the adversary, and consider ``quantization" the process of identifying the sensor output with an associated event. The local agent conveys the event realizations to its remote counterpart by losslessly encoding them into binary packets. In general, the length of the adversary's packets can vary over time. We allow the local agent to choose to ``not communicate" by sending an ``empty packet" $\emptyset$. If the local agent considers one event to be ``uninteresting" it might choose not to communicate when this event occurs. 

Let $\rvec{c}_{t}\in\{0,1\}^{*}$ denote the (possibly empty) packet, or \textit{codeword}, transmitted by the local agent to his remote counterpart at time $t$. We assume that the codewords $\rvec{c}_{0:T}$ are used to lossesly convey the sequence of quantizations $\rvec{q}_{0:T}$ to the remote site. We assume that the quantizations are defined on a countable space; without further loss of generality we assume that $\rvec{q}_{t}\in\mathbb{N}_{0}$. Let $Q:\Omega\rightarrow \mathbb{N}_{0}$ be the adversary's choice of quantization and encoding function so that \begin{align}\label{eq:quantization}
    \rvec{q}_{t} = Q\left(\rvec{y}_{t}\right). 
\end{align} We make the technical assumption that $Q$ is measurable. We assume that the quantization function is static ($Q$ depends neither on $t$ nor on previously transmitted packets). This is a simplifying assumption, and is reasonable when the source sequence $\{\rvec{y}_{t}\}$ is \textit{assumed to be} IID. We discuss generalizations in Section \ref{sec:conclusion}. The amount of data the remote site must process is equal to the amount of data the local agent transmits. The adversary's information processing cost is assumed to be proportional to the packet lengths, $\ell(\rvec{c}_{t})$. Say the time-horizon required for the remote site's estimation task is  $\s{T}+1$;
 the adversary's expected information processing cost is then 
\begin{align}\label{eq:commcost}
   \sum_{i=0}^{\s{T}} \mathbb{E}\left[ \ell(\rvec{c}_{t}) \right].
\end{align} The design of the encoding function $Q$, and the method by which the $\rvec{q}_{t}$ are encoded into the $\rvec{c}_{t}$ are assumed to be under the adversary's control. There are several methods the adversary can use to losslessly encode the $\rvec{q}_{t}$ into the codewords $\rvec{c}_{t}$. We discuss these methods after a few more definitions and stated assumptions.

We will assume that the $\{\rvec{y}_{t}\}$ are selected by the defenders. We associate $\{\rvec{y}_{t}\}$ with a \textit{nominal distribution}; based on, e.g., the physical configuration of the controlled plant. We assume that the adversary believes the  $\{\rvec{y}_{t}\}$ to be drawn from the nominal distribution, which we associate with the measure $\mu_{\rvec{y}}^{\mathrm{{N}}}$. Rationally inattentive adversaries will incorporate an understanding of $\mu_{\rvec{y}}^{\mathrm{{N}}}$ as well as their task goals into the design of their quantization function; this is the essence of data compression. We will assume that the nominal distribution of $\{\rvec{y}_{t}\}$ is such that the sequence is stationary with IID components, and will (abusively) overload the symbol $\mu_{\rvec{y}}^{\mathrm{{N}}}$ to refer to the marginal distribution. It will be useful to define a set of ``nominal" sensor readings  $\{\rvec{y}^{\mathrm{{N}}}_{t}\}\sim \mu_{\rvec{y}}^{\mathrm{{N}}}$. Under our initial assumptions and notation, if $\mathcal{E}_{i}\in \sigma(\Omega)$ then, e.g. $\mathbb{P}\left[ \cap_{i=0}^{T-1} \rvec{y}^{\mathrm{{N}}}_{i}\in\mathcal{E}_{i}   \right] = \prod_{i=0}^{T-1}\mu_{\rvec{y}}^{\mathrm{{N}}}\left[\mathcal{E}_{i}\right]$. We assume that the defender can select the sensor feeds from an arbitrary probability distribution so that  $\{\rvec{y}_{t}\} \sim \mu_{\rvec{y}}^{\mathrm{D}}$.
We now formulate an optimization for the defender to optimally select
$\mu_{\rvec{y}}^{\mathrm{D}}$. For our optimization, we show that $\mu_{\rvec{y}}^{\mathrm{D}}$ can be selected so that $\{\rvec{y}_{t}\}$ is a stationary, IID process without loss of generality.  

We will critically assume that the defenders know (or have black box access to) the adversary's choice of quantization mapping $Q$, and that they can use this knowledge to design a spoofing distribution $\mu_{\rvec{y}}^{\mathrm{D}}$. While this is a non-traditional notion of CTI, we propose that the set of quantization mappings that an adversary might reasonably use is deeply connected to both their \textit{goals} and \textit{beliefs} about the underlying system. An adversary that \textit{rationally inattentive}  will design their quantizer to minimize the expected information processing cost (\ref{eq:commcost}) subject to a constraint on the adversary's task performance at the remote site. We assume that the adversary believes that the honey sensor outputs are actually drawn from the nominal model. The nominal model could correspond to a ``real" sensor associated with the physical plant, or the adversary's belief could be otherwise established by baiting \cite{flamholz2019baiting}. 
We conjecture that, for a rationally inattentive adversary, the set of quantizers that might be reasonably used can be parameterized by the nominal model and the adversary's task goals.

The honey sensor outputs are entirely controlled by the defenders and can be selected arbitrarily. Our engagement technique will exploit this capability, and the defender's knowledge of the adversary's $Q$, to increase the remote site's information consumption. This is accomplished by producing sensor outputs that induce longer codewords. At the same time, we will control for the fact that the adversary is suspicious, and, upon observing atypical sensor outputs, might withdraw or pivot.  We now describe this mathematically. 

We will now develop a model for the codeword lengths in (\ref{eq:commcost}). The defender's choice of $\mu_{\rvec{y}}^{\mathrm{D}}$ and the adversary's quantizer $Q$ induces a distribution over $\{\rvec{q}_{t}\}$. Define the set-valued function $Q^{-1}:\mathbb{N}_{0}\rightarrow\sigma(\Omega)$ via $Q$'s inverse images so that
\begin{align}
    Q^{-1}(z) = \{y \in \Omega : Q(y) = z \}.
\end{align} 
In other words, $Q^{-1}(z)$ is the ``cell" of values in $\Omega$ that, when quantized, map to $z$. The \textit{entropy} of $\rvec{q}_{t}$ is then
\begin{IEEEeqnarray}{rCl}
    H(\rvec{q}_{t}) &=& -\sum_{z\in\mathbb{N}_{0}} \mathbb{P}\left[\rvec{q}_{t}=z\right]\log_{2}\left(\mathbb{P}\left[\rvec{q}_{t}=z\right]\right). 
\end{IEEEeqnarray} Making the dependence on $\mu_{\rvec{y}}^{\mathrm{D}}$ explicit, then
\begin{multline}
H(\rvec{q}_{t})= -\sum_{z\in\mathbb{N}_{0}} \mu_{\rvec{y}}^{\mathrm{D}}\left[ Q^{-1}(z)\right]\log_{2}\left(\mu_{\rvec{y}}^{\mathrm{D}}\left[ Q^{-1}(z)\right]\right) \nonumber
\end{multline} This formula is readily extended to the entropy of the \textit{sequence} $\rvec{q}_{0:T}$, i.e. $H(\rvec{q}_{0:T})$. We will argue that in our present setting we can justifiably assume that $H(\rvec{q}_{t})\approx \frac{1}{T}\sum_{i=1}^{T}\mathbb{E}[\ell(\rvec{c}_{t})]$. 

We first consider the case where the adversary uses a causal, zero delay encoder. In this case, the local agent uses an injective encoding function $E:\mathbb{N}_{0}\rightarrow\{0,1\}^{*}$ to produce the packet $\rvec{c}_{t}={E}(\rvec{q}_{t})$, and the adversary produces the inverse mapping at the remote site to recover the quantization. This sort of encoding is used, for example, in real-time tracking tasks. Let $\psi: \mathbb{R}^{+}\rightarrow \mathbb{R}^{+}$ via $\psi(x) =  x+(1+x)\log_2(1+x)- x\log_2(x)$. It can be shown that
\cite{verduVariableLength}
\begin{align}\label{eq:entropylb}
   \psi^{-1}(H(\rvec{q}_{t}))  \le \mathbb{E}\left[ \ell(\rvec{c}_{t}) \right].
\end{align} Note that $\psi^{-1}$ is increasing. If the local agent must convey the $\rvec{q}_{t}$ to the remote site in real time (before time $t+1$), this lower bounds the number of bits that the remote site receives for every $t$. The lower bound is \textit{tight} in that there exists an injective encoding function $\overline{E}:\{0,1\}^{*}\rightarrow\{0,1\}^{*}$ such that \cite{verduVariableLength,WYNER1972176}
\begin{align}\label{eq:entropyub}
 \mathbb{E}\left[ \ell(\overline{E}(\rvec{q}_{t})) \right] \le H(\rvec{q}_{t}).
\end{align}  The local agent could also convey the sequence $\rvec{q}_{0:T}$ to the remote station using a \textit{block code}. In this case, all the $\rvec{q}_{0:T}$ are encoded into a single codeword. This would be reasonable if the adversary is \textit{delay tolerant}, in that they can wait until time $t$ to receive all of $\rvec{q}_{0:t}$. Formally, if $E_{t}:(\mathbb{N}_{0})^{t}\rightarrow \{0,1\}^{*}$ is an injective block encoding function, we have, for any $t$ that  $\psi^{-1}(H(\rvec{q}_{0:t}))  \le \mathbb{E}\left[ \ell\left(E_{t}(\rvec{q}_{0:t})\right) \right].$ Note that if the $\rvec{q}_{t}$ are IID, then $H(\rvec{q}_{0:t}) = \sum_{k=0}^{t}H(\rvec{q}_{t})$. As in (\ref{eq:entropyub}) there is a matching upper bound, e.g. there exists an encoder $E_{t}$ achieving  $\mathbb{E}\left[ \ell(E_{t}(\rvec{q}_{0:t})) \right] \le H(\rvec{q}_{0:t})$. 

Designing the encodings $E$ and $E_{t}$ above generally requires knowledge of the underlying distribution of $\rvec{q}_{t}$. In our case the distribution of  $\rvec{q}_{t}$ is the product of the defender's deceptive spoofing, and thus is not immediately accessible to the adversary. Despite this, an adversary could use online universal source coding techniques to have approximately
\begin{align}
    \frac{1}{t}\mathbb{E}\left[ \ell\left(E_{t-1}(\rvec{q}_{0:t-1})\right) \right] \approx \frac{1}{t}H(\rvec{q}_{0:t-1})
\end{align} in the case of block coding, and analagously $\frac{1}{t}\sum_{i=0}^{t-1}\mathbb{E}\left[ \ell\left(E_{i}(\rvec{q}_{i})\right) \right] \lessapprox    \frac{1}{t}H(\rvec{q}_{0:t-1})+1$
 in the case that the adversary requires a zero-delay encoding. See, e.g. \cite{lz} for the case when $\{\rvec{q}_{t}\}$ is a stationary ergodic process on a finite alphabet, and \cite{frenchExp} for IID processes on countably infinite alphabets. 
The goal of the defender/spoofer is to increase the total information consumption of the adversary; we thus propose to design $\mu_{\rvec{y}}^{\mathrm{D}}$ to \textit{maximize} the entropy $H(\rvec{q}_{0:t})$. The general, a rationally inattentive adversary will transmit shorter codewords upon observing relatively more-likely (quantized) sensor outputs. Thus, increasing the adversary's information processing cost (\ref{eq:commcost}) via sensor spoofing will generally require the sensor's outputs to appear to be ``less likely" than they otherwise would be. We thus propose to control for the adversary's ability to detect whether or not the defenders are spoofing. We will take a conservative approach (generous to the adversary) that assumes the local agent performs optimal hypothesis testing to detect defender spoofing. An analagous approach was used in \cite{karabag2022deception,patil2023simulator}.  We make the technical assumption that the defender's choice of $\mu_{\rvec{y}}^{\mathrm{D}}$ has  $\mu_{\rvec{y}}^{\mathrm{D}}\ll \mu_{\rvec{y}}^{\mathrm{{N}}}$.\footnote{This is a reasonable restriction; no event that has positive probability under spoofing can have zero probability of occurring under the nominal distribution.  }

Assume that after $k$ timesteps, the adversary performs a simple Neyman-Pearson binary hypothesis test comparing the spoofed distribution with the nominal distribution, e.g.  the local agent computes the test statistic \begin{align}\label{eq:teststat}
    \rs{\gamma} =  \log_{2}\left(\frac{d\mu_{\rvec{y}}^{\mathrm{D}}}{d\mu_{\rvec{y}}^{\mathrm{{N}}}}(\rvec{y}_{0:k})\right),
\end{align} and concludes that he is being spoofed if $\rs{\gamma}\ge \omega$, where $\omega$ is a threshold determined by the local agent's tolerance for false alarms/missed detections. If $\Sigma = \mathbb{R}^{n}$ and both ${\rvec{y}}^{\mathrm{D}}$ and ${\rvec{y}}^{\mathrm{{N}}}$ are continuous random variables with PDFs $f_{\rvec{y}_{0:k}}^{\mathrm{D}}$ and $f_{\rvec{y}_{0:k}}^{\mathrm{{N}}}$ then the RN derivative in (\ref{eq:teststat}) can be replaced by the more-familiar likelihood ratio, e.g. $\rs{\gamma} =  \log_{2}\left(f_{\rvec{y}}^{\mathrm{D}}(\rvec{y}_{0:k})/f_{\rvec{y}}^{\mathrm{N}}(\rvec{y}_{0:k})\right)$.
 The expected value of the test statistic (given that $\rvec{y}_{0:k}\sim \mu_{\rvec{y}}^{\mathrm{D}}$) is exactly the KL divergence/relative entropy between the spoofed and nominal distributions of the sensor readings,
\begin{align}\label{eq:kl_test_stat}
    \mathbb{E}_{\rvec{y}_{0:k}\sim \mu_{\rvec{y}}^{\mathrm{D}}}\left[    \rs{\gamma} \right] = D_{\mathrm{KL}}\left(\mu_{\rvec{y}_{0:k}}^{\mathrm{D}}||\mu_{\rvec{y}_{0:k}}^{\mathrm{{N}}} \right). 
\end{align} We propose that from the defender's perspective it is desirable to design $\mu_{\rvec{y}}^{\mathrm{D}}$ to minimize this divergence. An analogous argument was proposed in \cite{patil2023simulator} in the context of deceptive control. As in \cite{patil2023simulator}, we can further motivate minimizing (\ref{eq:kl_test_stat}) via the Bretagnolle–Huber (BH) inequality. Let $A_{k}:\Omega^{k+1}\rightarrow [0,1]$ be the local agent's spoof detector at time $k$, so that the adversary will detect a spoof if $A_{k}(y_{0:k}) = 1$. By the BH inequality, the sum of the probabilities of false alarm and missed detection are bounded from below via
\begin{align}
    \mathbb{P}[A_{k}(\rvec{y}^{\mathrm{{N}}}_{0:k})=1] +         \mathbb{P}[A_{k}(\rvec{y}_{0:k})=0] \ge2^{-D_{\mathrm{KL}}(\mu_{\rvec{y}_{0:k}}^{\mathrm{D}}||\mu_{\rvec{y}_{0:k}}^{\mathrm{{N}}})-1}.\nonumber
\end{align} Thus, the KL divergence provides a lower bound on the detector's sum probabilities of false alarm and missed detection. Minimizing the divergence maximizes this bound, and as is desirable from the defender's perspective.

We design $\mu_{\rvec{y}}^{\mathrm{D}}$ via optimizing over probability measures from which we can draw $\rvec{y}_{0:T}$
\begin{align}\label{eq:initial_optimization}
\mu_{\rvec{y}}^{\mathrm{D}} =
 \underset{\substack{\mu\in \mathcal{P}(\sigma(\Omega)^{T+1})\text{, }\mu \ll \mu_{\rvec{y}}^{\mathrm{{N}}}\\ \rvec{y}_{0:T} \sim \mu\text{, } \rvec{q}_{t}=Q(\rvec{y}_{t})} }{\arg\max}\text{ }H(\rvec{q}_{0:T})-\lambda D_{\mathrm{KL}}(\mu||\mu_{\rvec{y}_{0:T}}^{\mathrm{{N}}} )\end{align} The intuition behind (\ref{eq:initial_optimization}) is to design a distribution of spoofed sensor outputs such that (1) the remote site has to process more data (e.g. to maximize $H(\rvec{q}_{0:T})$) while (2) controlling for the possibility that a suspicious local agent could halt transmissions if a spoof is detected (incentivizing minimization of $D_{\mathrm{KL}}(\mu_{\rvec{y}_{0:T}}||\mu_{\rvec{y}_{0:T}}^{\mathrm{{N}}} )$). The parameter ${\lambda}>0$ controls for the level of adversary suspicion. To solve (\ref{eq:initial_optimization}), the defender must know the adversary's quantizer design $\rvec{Q}$.

Given our assumption that the $\rvec{y}^{\mathrm{{N}}}_{t}$ are IID,  By Jensen's inequality, it can be seen that $D_{\mathrm{KL}}(\mu_{\rvec{y}_{0:T}}^{\mathrm{D}}||\mu_{\rvec{y}_{0:T}}^{\mathrm{{N}}})$ is minimized when $\mu_{\rvec{y}}^{\mathrm{D}}$ is such that $\rvec{y}_{0},\dots \rvec{y}_{T} $ are independent; i.e. for any choice of joint spoofing distribution $\mu_{\rvec{y}}^{\mathrm{D}}$ and $k$
\begin{align}
    \sum_{i=0}^{k}D_{\mathrm{KL}}\left(\mu_{\rvec{y}_{i}}^{\mathrm{D}}|| \mu_{\rvec{y}_{i}}^{\mathrm{{N}}}\right) \le D_{\mathrm{KL}}\left(\mu_{\rvec{y}_{0:k}}^{\mathrm{D}}|| \mu_{\rvec{y}_{0:k}}^{\mathrm{{N}}}\right).  
\end{align}   Likewise, we have the entropy $H(\rvec{q}_{0:k})$ is maximized when the $\rvec{q}_{t}$ are mutually independent. In this case, we have that
\begin{align}
    H(\rvec{q}_{0:k})\le \sum_{i=0}^{k}H(\rvec{q}_{i}).
\end{align} Note that if the $\{\rvec{y}_{i}\}$ are mutually independent, by (\ref{eq:quantization}) the  $\{\rvec{q}_{i}\}$ are mutually independent.  Thus without loss of generality we can restrict the optimization in (\ref{eq:initial_optimization}) to be over probability measures where the $\{\rvec{y}_{i}\}$ are mutually independent (replacing a $\mu$ with a ``mutually independent" version with equivalent marginals cannot decrease the objective). Thus we can replace $(\ref{eq:initial_optimization})$ with a ``single letterized" version
\begin{align}\label{eq:initial_optimization_single_letter}
\mu_{\rvec{y}_{i}}^{\mathrm{D}} =
 \underset{\substack{\mu \in \mathcal{P}(\mathcal{A})\text{, }\mu \ll \mu_{\rvec{y}}^{\mathrm{{N}}}\\ \rvec{y}_{i} \sim \mu} }{\arg\max}\text{ }H(Q(\rvec{y}_{i}))-\lambda D_{\mathrm{KL}}(\mu ||\mu_{\rvec{y}_{i}}^{\mathrm{{N}}} )\end{align}  which does not depend on $i$ since the $\rvec{y}_{i}^{\mathrm{{N}}}$ are IID. In the case that both the $\{\rvec{y}_{i}^{\mathrm{{N}}}\}$ and $\{\rvec{y}_{i}^{\mathrm{D}}\}$ are IID, the optimal way for the local agent to make a decision to detect ``spoofing" (in the sense of a sequential statistical test with minimum-time-to-decision for fixed probabilities of false alarm and missed detentions) is to perform a sequential likelihood ratio test \cite{wald1948sequential}. In this case, the expected time-to-decision is proportional to $1/D_{\mathrm{KL}}(\mu_{\rvec{y}_{i}}^{\mathrm{D}} ||\mu_{\rvec{y}_{i}}^{\mathrm{{N}}})$. We are thus justified in asserting that choosing a large  $\lambda$  in the optimization (\ref{eq:initial_optimization_single_letter}) increases the time an adversary will take to stop early. 
 
 In the next section, we show that (\ref{eq:initial_optimization_single_letter}) has a solution in certain relevant settings. We will discuss the case where the defender is uncertain as to the design of $Q$ in Section \ref{sec:numerical}.
 \section{Solutions to the optimization (\ref{eq:initial_optimization_single_letter})}\label{sec:optimization}
  It will useful to write entropy as an explicit function of the underlying probability measure; if $\rvec{y}\sim \mu$ we will write
\begin{align}
   H(\mu\circ Q^{-1}) =   H(Q(\rvec{y})).
\end{align} In measure theoretic terms, $H(Q(\rvec{y}))$ is the entropy of the pushforward of $\mu$ by the map $Q$. This notation makes explicit the dependence of the entropies on $\mu$. Consider the Lagrangian 
 \begin{multline}
     L(\mu,\phi) =  H(\mu\circ Q^{-1})-\lambda D_{\mathrm{KL}}(\mu ||\mu_{\rvec{y}_{i}}^{\mathrm{{N}}})+\phi(\int_{\Omega} d\mu-1),\nonumber
 \end{multline}  and the relaxation of (\ref{eq:initial_optimization_single_letter})
 \begin{align}\label{eq:lagrangeoptimization}
     \max_{\mu: \text{ a positive measure on $\sigma(\Omega)$} }L(\mu,\phi),
 \end{align} note that for any $\phi\in\mathbb{R}$, the optimization in (\ref{eq:lagrangeoptimization}) is convex (maximization of a concave functional subject to convex constraints). By definition the KL divergence term is $-\infty$ unless $\mu \ll \mu^{\mathrm{{N}}}$. We can seek a solution by computing a stationary point of (\ref{eq:lagrangeoptimization}); let $\tau \in \mathbb{R}^{+}$ and $\xi$ be a bounded signed measure on $\sigma(\Omega)$. If $\mu^{*}$ is an optimum of (\ref{eq:lagrangeoptimization}) it must be that $\mu^{*}$ is a positive measure with 
 $\mu^{*} \ll \mu^{\mathrm{{N}}}$ and that $\ddt L(\mu^{*}+\tau \xi,\phi)|_{\tau = 0} = 0$
 for all $\xi$.
If $Q:\Omega\rightarrow\{0,1\}^{*}$ then
\begin{multline}
           \ddt H\left((\mu^{*}+\tau\xi)\circ Q^{-1}\right)|_{\tau = 0} =\\ -\int_{y}  \left(1+\sum_{z}\mathbbm{1}_{{y} \in Q^{-1}(z)} \log\left(\mu^{*}[Q^{-1}(z)]\right)\right) d\xi(y),
 \end{multline} meanwhile
 \begin{multline}
           \ddt D_{\mathrm{KL}}\left((\mu^{*}+\tau\xi) || \mu^{\mathrm{{N}}} \right)|_{\tau = 0} =\\ \int \left(1+ \log\left(\frac{d\mu^{*}}{d\mu_{0}}(y)\right) \right) d\xi(y),
 \end{multline}  and $\ddt \phi(\int_{\Omega} d(\mu+\tau\xi)-1) =     \phi\int_{\Omega} d\xi$. Substituting these into  $\ddt L(\mu^{*}+\tau \xi,\phi)|_{\tau = 0}=0$ and  recognizing that the relation must hold for all $\xi$ gives that, for all $y\in {Q}^{-1}(z)$,
\begin{align}
    \log(\frac{d\mu^{*}}{d\mu_{0}}(y))=\frac{ -\log\left(\mu^{*}[Q^{-1}(z)]\right)}{\lambda} +\frac{\phi-1}{\lambda}-1.
\label{eq:set_to_zero2}
\end{align}
Thus, $ \log(\frac{d\mu^{*}}{d\mu_{0}}(y))$ is constant over sets of $\Omega$ falling within the same quantizer cell. This suggests that $\mu^{*}$ resembles
\begin{align}\label{eq:form}
    \mu^{*}[\mathcal{E}] = \int_{y\in\mathcal{E}}\sum_{z\in\mathbb{N}_{0}} \mathbb{1}_{y\in{Q}^{-1}(z)}r_{z}d\mu^{\mathrm{N}}(y)
\end{align} where the $r_{z}$ are positive constants such that $\mu^{*}$ is a probability measure. Substituting (\ref{eq:form}) into (\ref{eq:set_to_zero2}) gives
\begin{align}\label{eq:req}
    r_{z} = \begin{cases} 
    0 \text{, when } \mu^{\mathrm{N}}[{Q}^{-1}(z)] =0,\\
        e^{\frac{\phi-1-\lambda}{\lambda+1}} /{\mu^{\mathrm{{N}}}[Q^{-1}(z)]^{\frac{1}{1+\lambda}}}\text{, otherwise. } 
    \end{cases}
\end{align} It remains to solve for $\phi$, which should be selected to ensure that (\ref{eq:form}) is a valid probability measure. We ignore this challenge for the moment, returning after the next paragraph.

Consider again the case where $\Sigma= \mathbb{R}^{n}$ and $\mu^{\mathrm{N}}$ is absolutely continuous with respect to Lebesgue measure. In this case,  $\rvec{y}^{\mathrm{N}}\sim \mu^{\mathrm{N}}$ is a continuous random variable, and
has a PDF $f_{\rvec{y}^{\mathrm{N}}}$. In this case, the form (\ref{eq:form}) and weights (\ref{eq:req}) suggest that a random variable drawn from the optimal spoofing distribution $\rvec{y}^{\mathrm{D}}\sim\mu^{*}$ has a PDF given by 
\begin{align}\label{eq:densityexample}
     f_{\rvec{y}^{\mathrm{D}}}(x) = e^{\frac{\phi-1-\lambda}{\lambda+1}} f_{\rvec{y}^{\mathrm{N}}}(x) \sum_{z\in\mathbb{N}_{0}}\frac{\mathbb{1}_{x \in Q^{-1}\left(z\right)}}{\mathbb{P}[Q(\rvec{y}^{\mathrm{N}}) = z ]^{\frac{1}{1+\lambda}}}.
\end{align} Within each quantizer cell (e.g. when $x\in Q^{-1}\left(z\right)$) the probability density of $\rvec{y}^{\mathrm{D}}$ looks like a scaled version of that of $\rvec{y}^{\mathrm{N}}$. The scaling for a particular cell increases the \textit{less likely} the nominal sensor output is to fall within the cell. In particular, (\ref{eq:densityexample}) gives
\begin{align}\label{eq:problem1}
    \mathbb{P}[Q(\rvec{y}^{\mathrm{D}}) = z ] = e^{\frac{\phi-1-\lambda}{\lambda+1}} \mathbb{P}[Q(\rvec{y}^{\mathrm{N}}) = z ]^{\frac{\lambda}{1+\lambda}}.
\end{align} ``Flattening" the probability distribution of the quantizer's outputs increases their entropy. If the quantizer $Q$ has a finite range and we set $\lambda=0$ (ignoring the risk of detection), the optimal spoofing strategy is uniform (maximum entropy) over the quantizer support. Furthermore, as $\lambda\rightarrow \infty$ (tolerating no detection risk) we recover $ f_{\rvec{y}^{\mathrm{D}}}(x)= f_{\rvec{y}^{\mathrm{N}}}(x)$. 

When a normalizing $\phi$ can be found so that (\ref{eq:problem1}) gives a valid probability measure, (\ref{eq:problem1}) solves (\ref{eq:initial_optimization_single_letter}) and thus (\ref{eq:initial_optimization}). Unfortunately, finding such a $\phi$ can be problematic when $Q$ has infinite support. Since $\mu^{*}$ is a probability measure, we must have that 
\begin{IEEEeqnarray}{rCl}
1 = \int_{\Omega} d\mu^{*}\label{eq:normalization}  &=&  \sum_{z \in\mathbb{N}_{0}} r_{z} \mu^{\mathrm{{N}}}[{Q}^{-1}(z)]\nonumber\\ &=&  \sum_{ z\in\mathbb{N}_{0} } e^{\frac{\phi-1-\lambda}{\lambda+1}}\mu^{\mathrm{{N}}}\left[
 Q^{-1}(z)\right]^{\frac{\lambda}{1+\lambda}}  .\label{eq:problem2}
\end{IEEEeqnarray} If (\ref{eq:problem2}) converges, then $\phi$ can be selected so that substituting  (\ref{eq:req}) into (\ref{eq:form}) makes $\mu^{*}$ a probability measure. This is not always possible, although it is in some relevant special cases. 

If $Q$ maps to a finite number of quantization points under the nominal distribution (e.g. the range of $Q$ is finite or, more generally, the random variable
$\rvec{q} = {Q}(\rvec{y}^{\mathrm{N}})$ has finite support), (\ref{eq:problem2}) is has a finite number of terms. A finitely supported $Q$ is required for a digital system with finite precision. One might expect that (\ref{eq:problem2}) would converges if the quantizer's ``nominal" output entropy has $ H(\mu^{\mathrm{N}}\circ Q^{-1}) < \infty$. This is required for the adversary to expect finite communication cost under ``normal" circumstances, without the defender spoofing. Unfortunately, this condition is not sufficient (e.g. take $\lambda = 2$ and choose $Q$ and $\mu^{N}$ so that $\mu^{\mathrm{{N}}}\left[
 Q^{-1}(z)\right]\propto 1/z^{2}$).  If the probability mass function of $\rvec{q} = Q(\rvec{y}^{\mathrm{N}})$ has an exponentially bounded tail (cf. \cite{frenchExp}), (\ref{eq:problem2}) converges for all $\lambda$. In the remainder of this work, we focus settings where (\ref{eq:problem2}) converges, leaving further analysis to future work. Section \ref{sec:numerical} develops intuition for when the quantizer has finite range.   
 
Assume a normalizing $\phi$ can be shown to exist, and that the nominal sensor readings are continuous random variables so that  (\ref{eq:densityexample}) applies. Assume that the defender can easily evaluate, and sample from, the nominal PDF $f_{\rvec{y}^{\mathrm{N}}}$, and that the defender has \textit{black box} access to the function $Q$ (e.g., the defender can, absent the adversary, query $Q$ with inputs and observe the outputs). Under these circumstances, a practical strategy for the defender would be to estimate the probability mass function $\mathbb{P}[Q(\rvec{y}^{\mathrm{N}}) = z ]$ and then to use standard rejection sampling to sample from the optimal spoofing distribution $f_{\rvec{y}^{\mathrm{D}}}$. The defender need not compute $\phi$.  Analogous arguments apply when the nominal sensor measurements are discrete. 
\section{Numerical example}\label{sec:numerical}
In this section we let $\Sigma = \mathbb{R}$ and take the nominal sensor outputs as standard normal ($f_{\rvec{y}^\mathrm{N}}(x) = \frac{1}{\sqrt{2\pi}}e^{-\frac{x^2}{2}}$). We assume the adversary's quantizer is binary with 
$Q_{\mathrm{A}}(x) = \mathbbm{1}_{x\in [1,2]}$. With two quantization points and a zero-delay encoding, the adversary optimally does not communicate when the more likely outcome occurs. Under the nominal distribution, the local agent should ``send a message" to the remote site when the quantizer's output is $1$, and otherwise send $\emptyset$ (no message).

\begin{figure}
    \centering
    \includegraphics[width=1.06\linewidth]{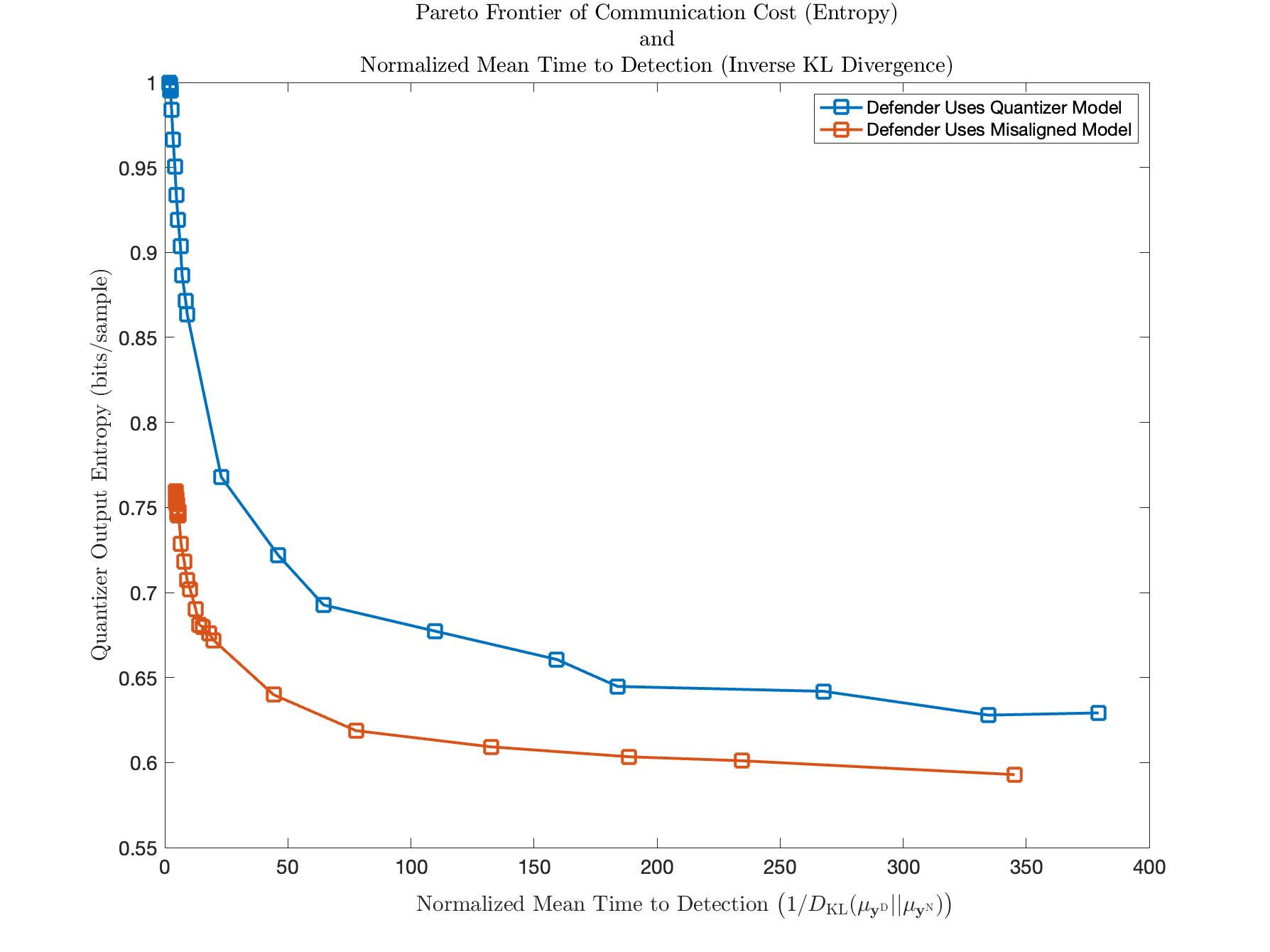}
    \vspace{-6mm}
    \caption{This plot illustrates the achieved adversary communication cost as a function of $1/D_{\mathrm{KL}}(\mu||\mu_{\rvec{y}_{0:T}}^{\mathrm{{N}}} )$. This inverse KL divergence is proportional to the ``time to detection" by the adversary under a sequential likelihood ratio test (cf. \cite{wald1948sequential}). The adversary is assumed to use the quantizer  $Q_{\mathrm{A}}$. The blue curve illustrates the case when the defender uses the correct quantizer to design the spoofing distribution, the red curve illustrates when the spoofer is designed for the wrong quantizer, $Q_{\mathrm{B}}(y)= Q_{\mathrm{A}}(y+.5)$.  The blue curve is the Pareto frontier of the entropy/divergence tradeoff computed by $\lambda$ in (\ref{eq:initial_optimization_single_letter}). The red curve's communication cost monotonically decreases only because $Q_{\mathrm{A}}^{-1}(1)$ and $Q_{\mathrm{B}}^{-1}(1)$ intersect; this will not occur in general. In either case, as the mean-time-to-detection tends to infinity the spoofing distribution tends to towards the nominal. The red and blue curves thus have horizontal asymptotes at the communication cost under the nominal distribution. 
 }
    \label{fig:pareto}
         \vspace{-7mm}
\end{figure}

In this case, one can numerically compute $\mu^{\mathrm{{N}}}[{Q}_{\mathrm{A}}^{-1}(z)]$ and it is straightforward to identify optimal spoofing PDF (up to a normalization) via (\ref{eq:densityexample}). We can then sample from the spoofed distribution using rejection sampling (using proposal samples from the nominal). Recall $\rvec{y}^{\mathrm{N}}\sim \mu^{N}$. The likelihood ratio of the spoofed to nominal distribution is simply
\begin{IEEEeqnarray}{rCl}
 \log\left( \dfrac{f_{\rvec{y}^{\mathrm{D}}}(y)}{f_{\rvec{y}^{\mathrm{N}}}(y)}\right) &=& \frac{\phi-1-\lambda-\log(\mathbb{P}[Q(\rvec{y}^{\mathrm{N}}) = Q(y) ])}{\lambda+1}.\nonumber
\end{IEEEeqnarray} 
\begin{figure}
    \vspace{-3mm}
    \centering
    \includegraphics[width=.87\linewidth]{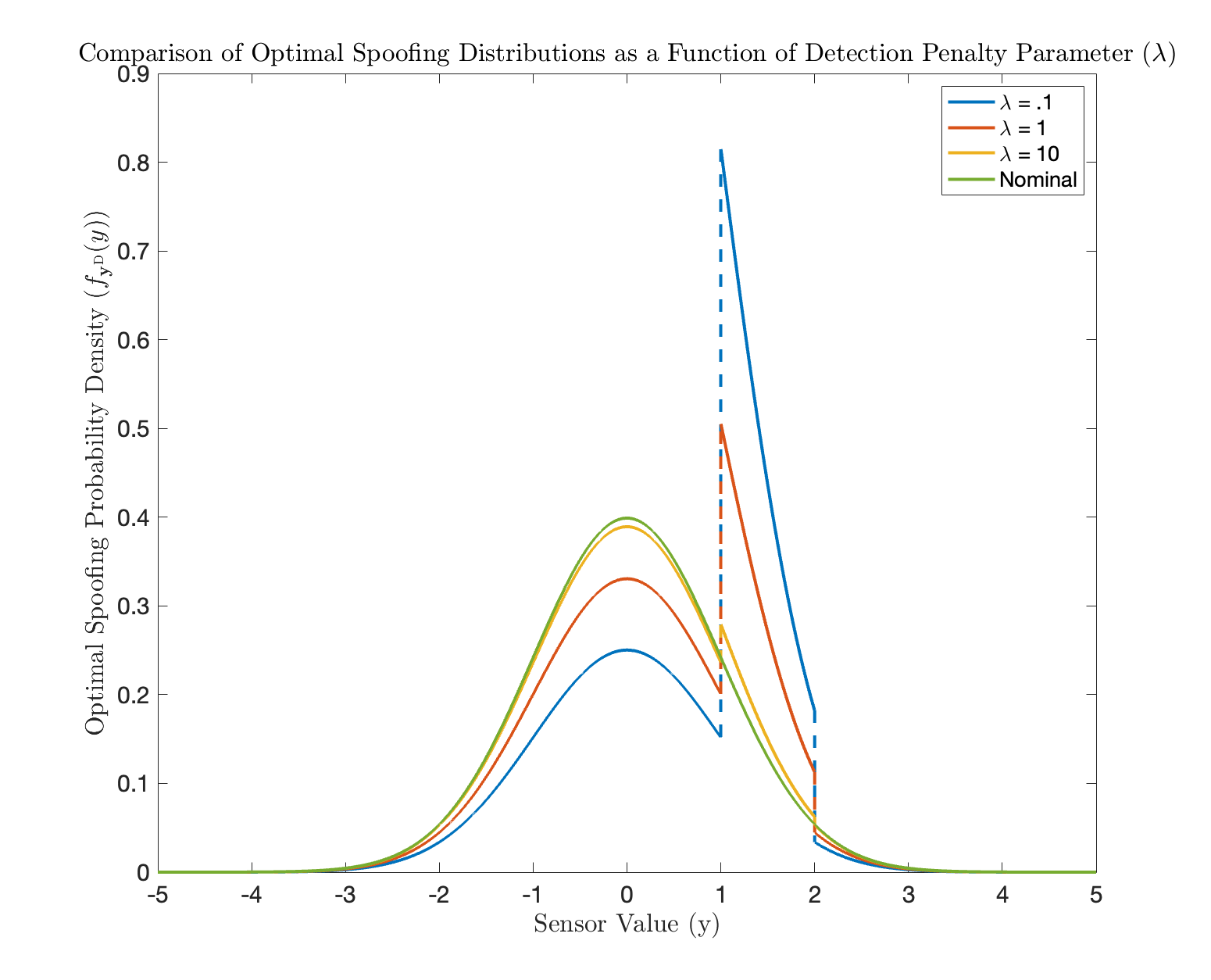}
    \vspace{-5mm}
    \caption{The PDF of the optimal spoofing distribution as a function of $\lambda$. Increasing $\lambda$ increases the ``cost" of a large KL divergence in the optimization (\ref{eq:initial_optimization_single_letter}). Convergence  to the nominal is observed with increasing $\lambda$. }
    \label{fig:pdf}
    \vspace{-2mm}
\end{figure} 
\begin{figure} 
    \centering
  \subfloat[\label{a}]{%
       \includegraphics[width=0.5\linewidth]{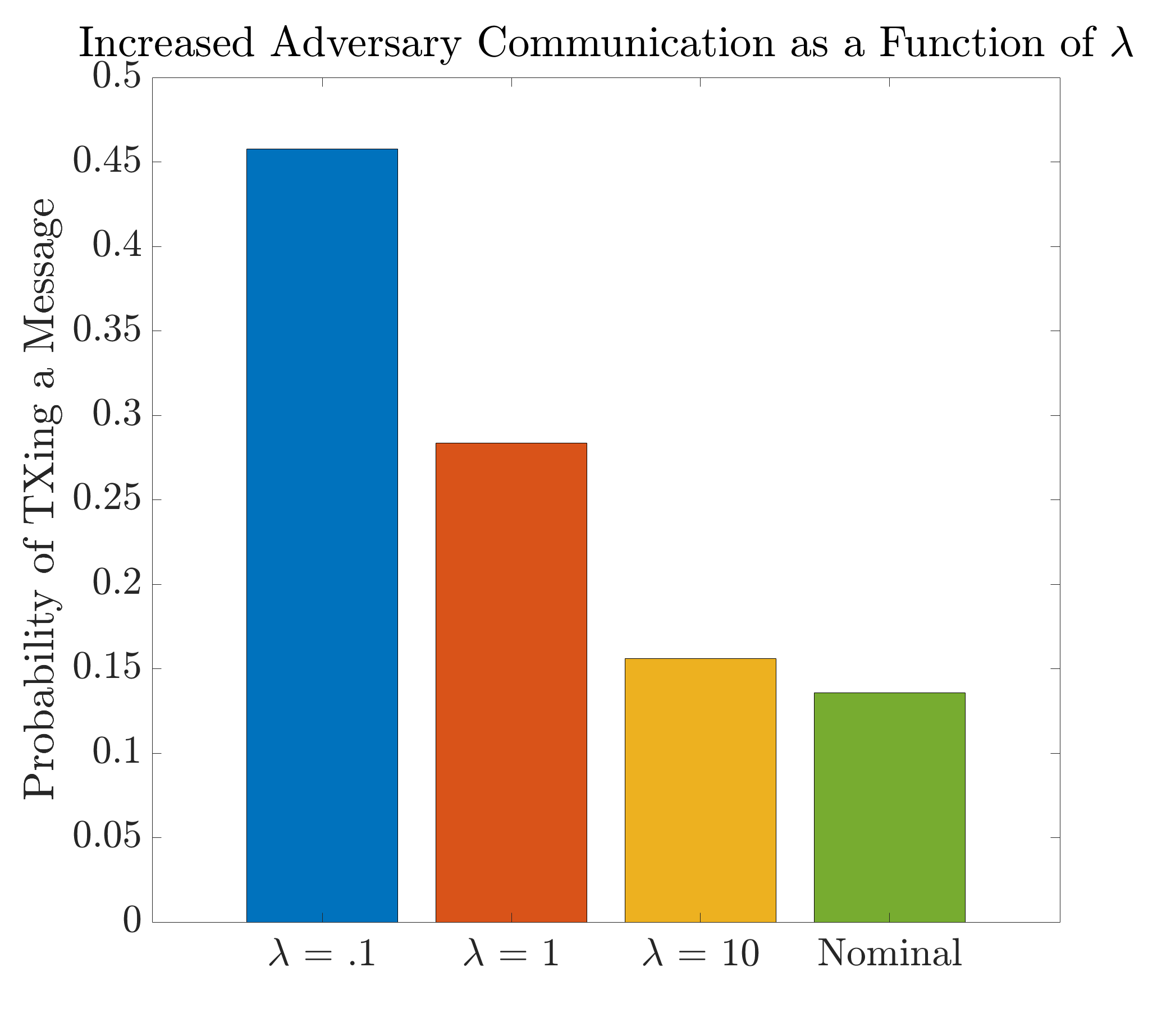}}
    \hfill
  \subfloat[\label{b}]{%
        \includegraphics[width=0.5\linewidth]{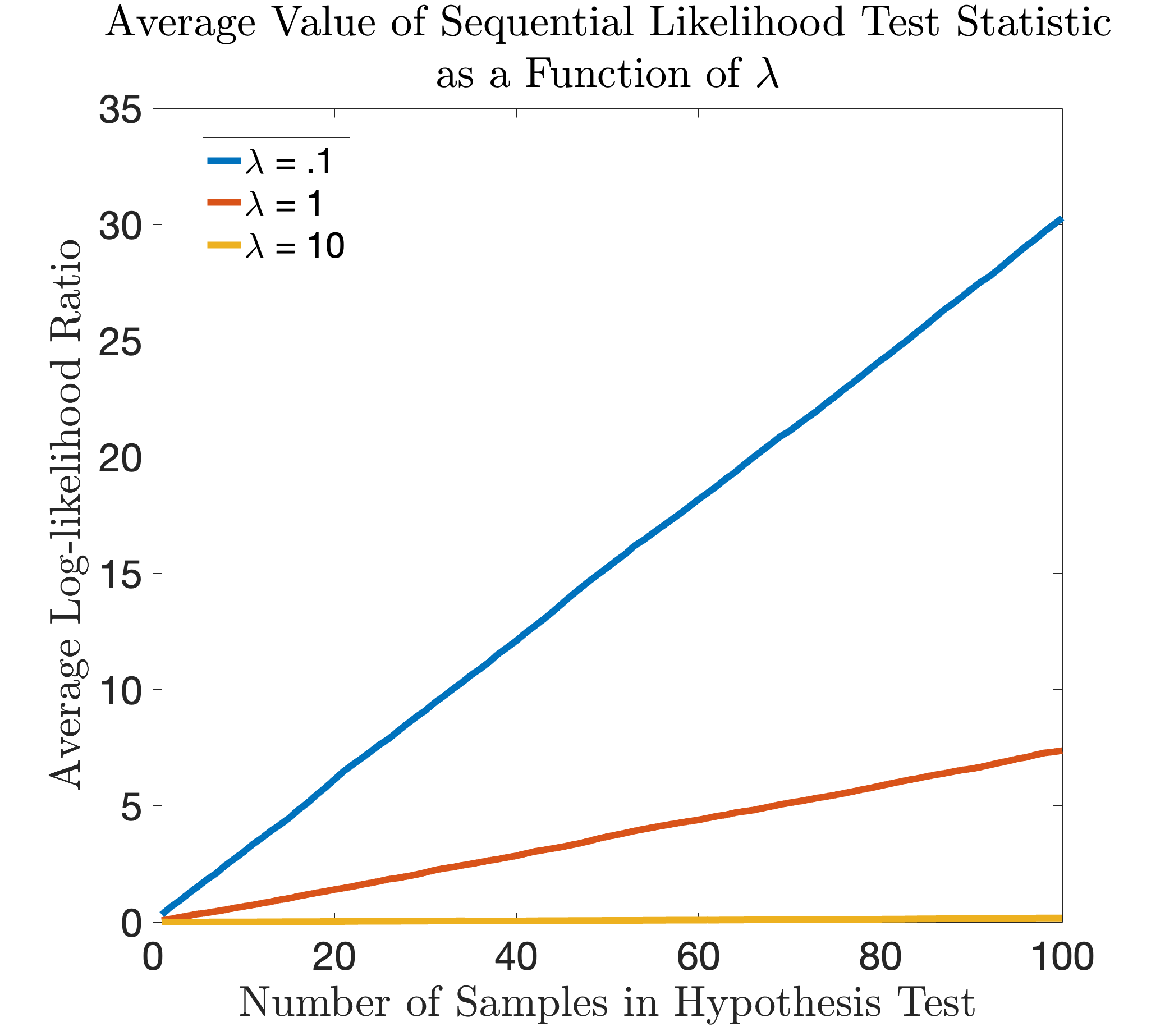}}
  \caption{In Figure \ref{a} the adversary's increased probability of communicating is computed for various choices of $\lambda$. Under our formulation, the adversary will never send a message more than 50$\%$ of the time; if such a situation were to occur, the adversary would instead send messages when the quantizer's input lies \textit{outside} of [1,2] (remaining silent otherwise). Figure \ref{b} illustrates the average value of the local agent's sequential likelihood ratio test statistic assuming the optimal test is used. The local agent would be expected to stop when the lines in \ref{b} cross a horizontal threshold. To maximize the total communication cost to the remote site, a defender should identify an operating point for $\lambda$ that accounts for the achieved communication cost per timestep, the ``stopping threshold", and the time horizon $T$. }
  \label{fig:double} 
\end{figure}
 We compare and discuss the remote site's increased expected communication cost (per sample) with the local agent's average likelihood ratio test statistic in Fig. \ref{fig:double}. A limiting assumption of our work is that it requires defender knowledge of the adversary's quantizer $Q$. In Fig. \ref{fig:pareto}, we consider a case where the defender computes the optimal spoofing distribution assuming the adversary uses $Q_{\mathrm{A}}$. We compare performance in the case that the adversary actually uses $Q_{\mathrm{A}}$ with the case when the adversary uses $Q_{\mathrm{B}}(y)= Q_{\mathrm{A}}(y+.5)$. Figure \ref{fig:pareto} illustrates the degradation in performance that results. 
\section{Conclusion and Future Work}\label{sec:conclusion}
While our numerical results were restricted to the case of a binary quantizer, we believe that this analysis might be of use when it comes to spoofing adversary systems that use an event-based communication paradigm. The operational communication cost incurred in an event-based communication architecture includes (theoretically absent, but practically necessary) \textit{overhead} (e.g., the bits that make up packet headers). This cost is incurred with every (nonempty) transmission. It might be reasonable to consider only a binary model of the adversary's quantizer capturing when nonempty transmissions occur. It would also be interesting to apply these techniques to fixed rate quantization and encoding schemes. Such schemes are often adaptive, and one could attempt to maximize the required amount of adaptation effort. 

Our proposed technique is limited by its assumption that the defender has access to the quantizer used by an adversary. This motivates future work along various avenues. One could consider a maximin version of (\ref{eq:initial_optimization}) where the defender's goal is instead to design a spoofing distribution that maximizes the minimum cost over all adversary quantizers within a given class. Alternatively, it would be useful to investigate learning optimal spoofing distributions online. Initially, one might consider a defender that selects, at each timestep, one of several candidate spoofing distributions with the same KL-divergence from the nominal. If the defender can observe the codeword lengths produced by the local agent, the situation resembles a mutli-armed bandit, for which there are optimized sampling techniques. The bandit model is especially suited to a notion of imperfect CTI where a defender is provided several candidate quantizers with initial confidences. Another limitation is our assumption that the nominal source distribution is IID (and the quantizer memoryless). In many cases our mathematical results can be generalized, but the computation required to design time-varying spoofers may be significant.

Choosing an operating point for $\lambda$ (equivalently, the allowable KL divergence between the spoofing and nominal distributions) for a given adversary profile is challenging. The best choice depends on the adversary's relative risk tolerances (missed spoof detections/false alarms). Inferring the optimal choice might require experimental investigation. Finally, it remains to understand the optimization when (\ref{eq:problem2}) diverges. While weak duality gives that the solution to (\ref{eq:lagrangeoptimization}) solves (\ref{eq:initial_optimization_single_letter}) when the solution is primal feasible;  establishing strong duality would provide the necessary insight.


%\section{}


\ifCLASSOPTIONcaptionsoff
  \newpage
\fi

\bibliographystyle{IEEEtran}
\vspace{-2mm}
\bibliography{IEEEabrv,references}
\clearpage

\end{document}